\documentclass[11pt]{amsart}

\usepackage{amsfonts,latexsym,array,delarray,amsthm,amssymb}
\usepackage{graphicx}

\theoremstyle{plain}
\newtheorem{thm}{Theorem}
\newtheorem{lemma}[thm]{Lemma}
\newtheorem{prop}[thm]{Proposition}
\newtheorem{cor}[thm]{Corollary}

\newtheorem{alg}[thm]{Algorithm}

\theoremstyle{definition}
\newtheorem{defn}[thm]{Definition}
\newtheorem{ex}[thm]{Example}

\theoremstyle{remark}

\begin{document}

\title[Reverse-engineering]{Reverse-engineering of polynomial dynamical systems}
\author[Jarrah]{Abdul Salam Jarrah}
\author[Laubenbacher]{Reinhard Laubenbacher}
\author[Stigler]{Brandilyn Stigler}
\author[Stillman]{Michael Stillman}

\address{Virginia Bioinformatics Institute, Virginia Tech, Blacksburg, VA 24061-0477, USA}
\email{ajarrah@vbi.vt.edu}

\address{Virginia Bioinformatics Institute, Virginia Tech, Blacksburg, VA 24061-0477, USA}
\email{reinhard@vbi.vt.edu}

\address{Mathematical Biosciences Institute, The Ohio State
University, Columbus, OH 43210}
\email{bstigler@mbi.osu.edu}

\address{Mathematics Department, Cornell University, Ithaca, NY 14853, USA}
\email{mike@math.cornell.edu}

\date{\today}



\thanks{The authors thank Miguel Col\'on-Velez for the
implementation of the statistical measures.  The first and second
authors were supported partially by NSF Grant DMS-0511441. The
second and third authors were supported partially by NIH Grant RO1
GM068947-01, a joint computational biology initiative between NIH and NSF.  The fourth
author was supported partially by NSF Grant DMS-0311806. }

\begin{abstract}
Multivariate polynomial dynamical systems over finite fields have
been studied in several contexts, including engineering and
mathematical biology.  An important problem is to construct models
of such systems from a partial specification of dynamic
properties, e.g., from a collection of state transition
measurements. Here, we consider static models, which are directed
graphs that represent the causal relationships between system
variables, so-called wiring diagrams.  This paper contains an
algorithm which computes all possible minimal wiring diagrams for
a given set of state transition measurements. The paper also
contains several statistical measures for model selection.  The
algorithm uses primary decomposition of monomial ideals as the
principal tool. An application to the reverse-engineering of a
gene regulatory network is included.  The algorithm and the
statistical measures are implemented in Macaulay2 and are
available from the authors.
\end{abstract}

\maketitle

\section{Introduction}
\label{intro}

A {\it polynomial dynamical system} (PDS) over a finite field $k$ is
a function
$$
f=(f_1,\ldots ,f_n): k^n\longrightarrow k^n,
$$
with the coordinate functions $f_i\in k[x_1, \ldots ,x_n]$.
Iteration of $f$ results in a time-discrete dynamical system. PDSs
are special cases of {\it finite dynamical systems}, which are
maps $X^n\longrightarrow X^n$ over arbitrary finite sets $X$. Such
systems arise in applications in engineering (see, e.g., \cite{C,
E, LB, ML1, ML2, MW, WM} as well as in biology \cite{A, K2, LS}).
They include in particular the class of Boolean networks
($k=\mathbb{F}_2$) and cellular automata, which are studied and
applied extensively in computer science, engineering, and life
science disciplines \cite{K1,S1,D, GO, H, S2}. The problem of
finding systems with specified dynamic properties has arisen in
several contexts, e.g., \cite{LS, MW}.

The main result in this paper is an algorithm that identifies
minimal sets of variables for which a model exists that explains
the data. That is, the output consists of all possible minimal
``wiring diagrams'' of dynamic networks that fit the data. This
paper also relates to the algorithm in \cite{K2}.  There, the
question is considered what can be inferred about a biological
network from a set of experiments.  The network is represented by
a mapping $f:X^n\longrightarrow X^n$, where $n$ is the number of
variables and $X$ is a finite set of state values for the
variables.  It is shown in \cite{K2} that the problem of finding a
minimal wiring diagram for the network from a set of observations
is NP-hard.  The author then describes a greedy algorithm that
finds a wiring diagram which is ``close" to minimal and as sparse
as possible in polynomial time.  The algorithm in this paper can
be interpreted as a deterministic method that finds ALL possible
minimal wiring diagrams.  This is accomplished by imposing an
algebraic structure on the set $\Lambda$ and using tools from
symbolic computation. The main idea is to construct a square-free
monomial ideal from the given observations, whose minimal primes
are generated by the desired minimal variable sets. This method is
quite effective for networks arising from biological systems, even
though the general problem is NP-hard.

\section{Reverse-engineering of Polynomial Dynamical Systems}
\label{rev-eng}

We first describe briefly the algorithm in \cite{LS} and how it can
be applied to the reverse-engineering of biochemical networks, in
order to create a context for the algorithm in the present paper.
Recent technological advances, such as DNA microarray chips, have
made it possible to make large system-level measurements of chemical
species in cell extracts, such as gene transcripts from large
numbers of genes.  It would be desirable to have efficient
computational methods to extract information about regulatory
interactions between genes from repeated measurements of gene
transcript concentrations.  The relatively high cost of gene chip
technology makes it infeasible to collect sufficient time-course
measurements to uniquely determine the network.  A further
complication is that the variance in gene chip data is still
significant, adding additional complexity to the problem.

Any approach to network reconstruction must in principle proceed in
three steps: (1) choose a model type, e.g., systems of differential
equations or Bayesian networks.  (2) Describe the space of models
that are consistent with the given data set.  (3) Choose a ``most
likely'' model that generated the given data set, based on
predefined selection criteria.  From this model, one can then
determine the structure of the network, or its dynamic
behavior.

In \cite{LS} the authors proposed as choice in Step (1) the class
of polynomial dynamical systems over finite fields. This choice
was motivated by several considerations.  Molecular data sets such
as DNA microarray measurements are still sufficiently noisy to
justify qualitative models that distinguish only finitely many
possible states.  Furthermore, biologists typically interpret such
data as giving a relatively small number of regulatory states,
e.g. fold-change over control measurements. Finally, discrete
models for biochemical networks have a long and successful
tradition, beginning with the work of S. Kauffman \cite{K1}, the
logical models of Snoussi and Thomas \cite{ST}, and, more
recently, Bayesian network models \cite{F}.

Once this model class is chosen, we need to address Step 2, the
description of the model space corresponding to a given data set.
To be precise, suppose we are considering $n$ variables $x_1,
\ldots , x_n$, e.g., gene transcript concentrations for $n$ genes.
Suppose that we have measured $m$ real-valued state transition
pairs $(\mathbf{s}_1,\mathbf{t}_1),\ldots
,(\mathbf{s}_m,\mathbf{t}_m)$ with
$$
\mathbf{s}_i=(s_{i1},\ldots , s_{in}),\mathbf{t}_i=(t_{i1}, \ldots
,t_{in}), i=1, \ldots ,m,
$$
where $s_{ij}, t_{ij}\in \mathbb R$. That is, if the system is in
state $\mathbf{s}_i$, then it transitions to state $\mathbf{t}_i$
at the next time step. The first task is to choose a suitable
finite field $k$ and associated discrete state transition pairs
${\mathbf s}_i, {\mathbf t}_i$ in $k^n$.  This is a difficult and
very important step.  Most existing clustering algorithms are not
suitable for this purpose, and a new algorithm is proposed in
\cite{DLM}. An admissible model then is a function
$f:k^n\longrightarrow k^n$ such that
\begin{equation}
\label{re} f({\mathbf s}_i)={\mathbf t_i}.
\end{equation}
Since any function $k^n\longrightarrow k$ can be represented as a
polynomial function \cite[p. 369]{LN}, the model space under
consideration is the collection of all PDSs $f = (f_1, \ldots, f_n)$
such that equation (\ref{re}) holds.

Note that, if $f$ and $g$ are two such models, then $f-g$ is a
polynomial function that vanishes identically on the given data set.
Hence we can compute the model space by computing a particular model
$f^0 = (f^0_1, \ldots, f^0_n)$,
 where the coordinate function $f^0_i \in k[x_1,\ldots ,x_n]$,
and the ideal $I$ of the variety given by the points ${\mathbf
s}_i$. This ideal of points can be computed efficiently by the
Buchberger-M\"oller algorithm (see \cite{R} for a description of
the algorithm). The model space is given by the tuple of cosets
$(f_1^0 + I, \ldots, f_n^0 + I)$. It is clearly sufficient to
solve the network reconstruction problem for each network node
separately.

Step (3) then consists of the selection of a ``most likely'' model
from the space $h + I$, where $h \in k[x_1, \ldots, x_n]$. In
\cite{LS}, it was proposed to choose the normal form (reduction)
of any particular model $h$ with respect to a Gr\"obner basis for
the ideal $I$. The motivation for this choice was that the model
should be minimal, in the sense that it should not contain any
terms that vanish identically on the data. Inclusion of such terms
may affect the inference of network structure by introducing
interactions in the network that are not supported by the data.
However, this choice is unsatisfactory for several reasons. In
particular, the normal form depends on the choice of a particular
term order to compute a Gr\"obner basis, and there is no canonical
choice for such a term order.  The algorithms in this paper are
partially motivated by an effort by the fourth author to make the
model selection criterion less dependent on the choice of term
order.

In the next section we describe an algorithm that finds all sets
of variables ${x_{i_1},\ldots ,x_{i_r}}$ for which
$$
k[x_{i_1},\ldots ,x_{i_r}]\cap (h+I)\neq\emptyset ,
$$
and such that the intersection is empty whenever one of the
variables is removed.  That is, this algorithm finds all minimal
sets of variables for which there exists a model consistent with
the given data.  One advantage is that this algorithm does not
depend on the choice of a term order.  In a later section we will
describe how it can be used for the purpose of model selection in
Step (3).

\section{Minimal variable sets via primary decomposition}
\label{algo}

We restrict our attention to the data for a single coordinate. That
is, we have the given data set
$$
({\mathbf s}_1, t_1),\ldots ,({\mathbf s}_m, t_m),
$$
where ${\mathbf s}_i\in k^n$ and $t_i\in k$ (notice that this is a
value in $k$ and not an $n$-tuple). We are interested in functions
$f\in k[x_1,\ldots ,x_n]$ such that $f({\mathbf s}_i)=t_i$. For
$a\in k$, let
$$
X_a=\{{\mathbf s}_i\mid t_i=a\},
$$
and $X=\{X_a\mid a\in k\}$. Write the model space $h+I$ for the
given data set as
$$
Y = \{f \in k[x_1, \ldots, x_n] \mid f({\mathbf p}) = a, \mbox{\ for
all ${\mathbf p} \in X_a,a \in k$}\}.
$$
We are interested in the elements $f$ of $Y$ which involve a
minimal number of the variables, in the sense that
there exists no $g \in Y$ such that the support of $g$ is properly
contained in the support of $f$.
To compute these minimal
variable sets, we encode them in a simplicial complex.

\begin{defn}
For $F \subset \{1,\ldots ,n\}$, let $R_F = k[x_i \mid i \not\in
F]$.  Let
$$
\Delta_X := \{ F \subset \{1,\ldots ,n\} \mid Y \cap R_F \neq \emptyset\}.
$$
\end{defn}

Notice that $\Delta_X$ is a simplicial complex, since if $G\subset
F$, then $R_F \subset R_G$.  We now associate a square-free
monomial ideal to $\Delta_X$.

\begin{defn}\label{maindef}
Given $X$ as above, let $M_X \subset k[x_1, \ldots, x_n]$ be the
square-free monomial ideal generated by
$$
 W =  \{ m({\mathbf p},{\mathbf q}) \mid {\mathbf p} \in X_a,
  {\mathbf q} \in X_b, \mbox{and\ } a\neq b\in k\},
$$
where
  $$ m({\mathbf p},{\mathbf q}) := \prod_{p_i \neq q_i} x_i.$$
\end{defn}

Note that the monomial $m({\mathbf p}, {\mathbf q})$ encode the
coordinates in which ${\mathbf p}$ and ${\mathbf q}$ differ. Note
that $M_X$ is the same as the face ideal for the Alexander dual of
$\Delta_X$.  The following fact is the key technical result
underlying the algorithm.

\begin{prop}
For a given subset $F \subset \{1,\ldots ,n\}$, $F \in \Delta_X$
if and only if the ideal $\langle x_i \mid i \not\in F\rangle$
contains the monomial ideal $M_X$.
\end{prop}

\begin{proof}
First, assume that $F\in \Delta_X$.  Then $Y \cap R_F \neq
\emptyset$.  Let ${\mathbf p}\in X_a, {\mathbf q}\in X_b$, with
$a\neq b$. Then there exists $f\in k[x_i \mid i \not\in F]$ such
that $ f({\mathbf p})= a$ and $f({\mathbf q})=b$.  This implies
that $\mathbf p$ and $\mathbf q$ must differ in some coordinate
$j\not\in F$. Hence, the monomial $m({\mathbf p},{\mathbf q})$
must contain $x_j$ as a factor, and is therefore contained in
$\langle x_i \mid i \not\in F\rangle$. This proves one direction.

For the other direction, assume that $M_X\subset \langle x_i \mid
i \not\in F\rangle$.  Then all $m({\mathbf p},{\mathbf q})$ are
contained in this ideal, which implies that ${\mathbf p}\in X_a$
and ${\mathbf q}\in X_b, a\neq b$, differ in a coordinate
$i\not\in F$. Let $f$ be a function which sends all ${\mathbf
p}\in X_a$ to $a$ for all $a$.  This function can be represented
as a polynomial $f$, and, since ${\mathbf p}\in X_a$ and ${\mathbf
q}\in X_b$ differ in coordinates $i\not\in F$, this polynomial can
be expressed using only variables $x_i, i\not\in F$.  Hence $f\in
Y \cap R_F$.  This completes the proof of the proposition.
\end{proof}

The following corollary is immediate.

\begin{cor}
The minimal subsets $F$ such that $Y \cap R_{F} \neq \emptyset$
are precisely the generating sets for the minimal primes in the
primary decomposition of the ideal $M_X$.
\end{cor}

This corollary forms the basis for Algorithm~\ref{minsetsalg}.

\medskip
\begin{alg}
\label{minsetsalg}
  \rm
(\textbf{Minimal Sets})

\begin{tabular}{l}
    \hline
    \textbf{Input}: \quad $\{({\mathbf s}_1,t_1),\ldots ,({\mathbf s}_m, t_m)\}$,
with ${\mathbf s}_i\in k^n, t_i\in k$\\
    \textbf{Output}: All minimal subsets $F\subset\{1, \ldots ,n\}$ such
that there exists a \\
\qquad \qquad \quad polynomial function $f\in k[\{x_i\mid i\not\in
F\}]$ with $f({\mathbf s}_i)=t_i$.\\ \hline
    \quad   {\bf Step 1.} Compute the ideal $M_X$. \\
    \quad   {\bf Step 2.} Compute the primary decomposition of $M_X$.\\
    \quad   {\bf Step 3.} Compute the generating sets of all minimal primes of
$M_X$. \\ \hline
\end{tabular}
\end{alg}
\smallskip

\begin{ex}
\label{ex} Consider the following set of state transition pairs,
with entries in the field $\mathbb F_5$ with five elements:
\begin{center}
$\mathbf s_1=((3, 0, 0, 0, 0),3),$\\
$\mathbf s_2=((0, 1, 2, 1, 4),1),$\\
$\mathbf s_3=((0, 1, 2, 1, 0),0),$\\
$\mathbf s_4=((0, 1, 2, 1, 1),0),$\\
$\mathbf s_5=((1, 1, 1, 1, 3),4).$
\end{center}
Then $X_0=\{\mathbf s_3,\mathbf s_4\}$, $X_1=\{\mathbf s_2\}$,
$X_2=\emptyset$, $X_3=\{\mathbf s_1\}$, and $X_4=\{\mathbf s_5\}$.
Recall that the ideal $M_X$ is generated by all monomials
$m(\mathbf{p},\mathbf{q})$ with ${\mathbf p} \in X_a$ and
${\mathbf q} \in X_b$ for every $a\neq b\in k$. For $a=1$ and
$b=3$, the monomial $m((0,1,2,1,4),(0,1,2,1,0))=x_5$ since the two
points differ in the fifth coordinate, whereas
$m((0,1,2,1,4),(1,1,1,1,3))=x_1x_3x_5$. The ideal $M_X$ is
generated by the monomials $\{x_1x_2x_3x_4, x_5\}$ and has
associated primes $ \langle x_1, x_5\rangle, \langle x_2,
x_5\rangle, \langle x_3, x_5\rangle, \langle x_4, x_5\rangle $.
The minimal sets of variables required to define a function for
the data given above are:
$$
\{ x_1, x_5\}, \{ x_2, x_5\}, \{ x_3, x_5\}, \{ x_4, x_5\}.
$$
\end{ex}

 We now discuss
 the  complexity of Algorithm \ref{minsetsalg}.

 \begin{lemma}
    Let $|k| = q$ and let $W$ be the generating set of the ideal $M_X$ as in Definition \ref{maindef}. Then
    \[
    |W| \leq \dfrac{1}{2}(1-\dfrac{1}{q})m^2.
    \]
 \end{lemma}
\begin{proof}
Suppose that $X = \bigcup_{l=1}^j X_{i_l}$, where $i_l \in k$, for all $l$.
Suppose $|X_{i_l}| = r_{i_l}$ for all $l$. Then
\[
|W| \leq \sum_{g,h =1,g\neq h}^j r_{i_g} r_{i_h} \leq {q \choose
2} (\dfrac{m}{q})^2 = \dfrac{1}{2}(1-\dfrac{1}{q})m^2.
\]
\end{proof}

There are two well-known methods for finding the primary
decomposition of a monomial ideal like $M_X$, and the most
efficient one uses the Alexander dual of $M_X$ \cite{HS}. This is
the method used in our implementation in Macaulay 2 \cite{M2} of
the algorithms presented in this paper. Using the notion of a
monomial tree in the Alexander dual approach, it is conjectured
\cite{M} that the irreducible primes of $M_X$ can be found with
order $O(|W|^2
\log(n))$.  

\section{Model selection}

The algorithm described in the previous section produces a
potentially very large number of possible models for a given data
set.  If additional information about the network is available,
e.g., the existence or absence of certain interactions, then this
can be used for model selection.  This is the case for the
application discussed in the next section.  In this section we
describe a small collection of statistical measures for model
selection in the case where no additional information about the
network is available.  Each of the measures has a different set of
underlying assumptions, ranging from a bias toward small sets to a
bias toward large sets containing variables that appear in many of
the minimal sets.  Which measure is appropriate to use depends on
the type of network under consideration.

Let $\{x_1,\ldots ,x_n\}$ be the set of variables and let
$$
\mathcal{F} = \{F_1,\ldots ,F_t \} \subset \mathcal P(\{x_1,\ldots
,x_n\}),
$$
be the output of Algorithm \ref{minsetsalg} for a given data set.
We will construct several statistical measures on this output that
allow the choice of one or more subsets/models with highest
probability.

For  $1 \leq  s \leq n$, let $Z_s$ be the number of sets $F_j$ in
$\mathcal{F}$ of length $s$, and for $x_i \in \{x_1,\ldots
,x_n\}$, let $W_i(s)$ be the number of sets $F_j$ in $\mathcal{F}$
of length $s$ such that $x_i \in F_j$. That is,
$$Z_s = |\{j \, : \, F_j \in \mathcal{F} \mbox{ and } |F_j| = s \}|
\, \mbox{ and } $$
$$W_i(s) = |\{ j \, : \, F_j \in \mathcal{F},
\, x_i \in F_j \mbox{ and } |F_j| = s \}|.$$
We propose three different methods to score each variable $x_i \in
\{x_1,\ldots ,x_n\}$. Let
\begin{eqnarray*}
    S_1(x_i) &=& \sum_{s=1}^n \dfrac{W_i(s)}{s \cdot Z_s}, \\
    S_2(x_i) &=& \sum_{s=1}^n \dfrac{W_i(s)}{s}, \\
    S_3(x_i) &=& \sum_{s=1}^n W_i(s).
\end{eqnarray*}
Next we propose two different methods to compute a score for each
set $F_j \in \mathcal{F}$. Define
\begin{eqnarray*}
T_1(F_j) &=& \prod_{x_i\in F_j}S(x_i), \\
T_2(F_j) &=& \dfrac{\sum_{x_i \in F_j} S(x_i)}{|F_j|},
\end{eqnarray*}
where $S(x_i)$ is the score of $x_i$ using any of the three
variable-scoring methods.
If we now normalize the set scores by dividing by
$D=\sum_jT_i(F_j), i=1,2$, then we obtain a probability
distribution on the set $\mathcal{F}$.

\medskip\noindent
{\bf Example.}  Let $n=6$, and let
\[
\mathcal F=\{F_1=\{x_1\}; \, F_2=\{x_2,x_3\}; \, F_3=\{x_2,x_4\};
\, F_4=\{x_3,x_5,x_6\}\}.
\]
Then
\begin{center}
\begin{tabular}{|l|c|c|c|} \hline
         &  $S_1$ & $S_2$  & $S_3$ \\ \hline
   $x_1$ & $1$ &  $1$  &   $ 1 $    \\ \hline
   $x_2$ & $\frac{1}{2}$ & $\frac{1}{2}$ & $2$\\ \hline
   $x_3$ & $\frac{7}{12}$ & $\frac{5}{6}$ & $2$\\ \hline
   $x_4$ & $\frac{1}{4}$ & $\frac{1}{2}$ & $1$ \\ \hline
   $x_5$ & $\frac{1}{3}$ & $\frac{1}{3}$ & $1$ \\ \hline
   $x_6$ & $\frac{1}{3}$ & $\frac{1}{3}$ & $1$ \\ \hline
\end{tabular}
\end{center}
The following table lists set scores using different combinations
of the scoring methods $S_i$ and $T_j$.
\begin{center}
\begin{tabular}{|l|c|c|c|c|c|c|} \hline
         &  $S_1,T_1$ & $S_1,T_2$  & $S_2,T_1$  & $S_2,T_2$ & $S_3,T_1$  & $S_3,T_2$ \\ \hline
   $F_1$ & $1$ &  $1$&  $1$&  $1$&  $1$&  $1$    \\ \hline
   $F_2$ & $\frac{7}{24}$&$\frac{13}{24}$ & $\frac{10}{24}$& $\frac{16}{24}$& $4$&  $2$   \\ \hline
   $F_3$ & $\frac{1}{8}$& $\frac{3}{8}$& $\frac{2}{8}$& $\frac{4}{8}$& $2$& $\frac{6}{8}$   \\ \hline
   $F_4$ & $\frac{7}{108}$& $\frac{45}{108}$& $\frac{10}{108}$& $\frac{54}{108}$& $2$&  $\frac{4}{3}$   \\ \hline
\end{tabular}
\end{center}

Using either $T_1$ or $T_2$ to score sets, we would choose $F_1$
if we use $S_1$ or $S_2$, and $F_2$ if we use $S_3$.

\bigskip

Note that using $S_1$ or $S_2$ to score variables does not always
pick the singleton sets as one might suspect from the example
above. Suppose we have the collection of sets
$$
\{x_1\},\{x_{13}\},
\{x_2,x_3\},\{x_2,x_4,x_5\},\{x_2,x_6,x_7\},\{x_2,x_8,x_9\},
\{x_2,x_{10},x_{11}, x_{12}\}.
$$
Then $T(\{x_2,x_3\})>T(\{x_1\})$, where $T$ is either $T_1$ or
$T_2$, using any of the variable-scoring methods above.

\medskip
\begin{alg}
\label{modelselectionalg}
  \rm
(\textbf{Model Selection})

\begin{tabular}{l}
    \hline
    \textbf{Input}:
    \quad A collection of subsets $F_1,\ldots ,F_t$ of $\{x_1, \ldots ,x_n\}$.\\
    \textbf{Output}:  Return the set(s) with highest probability score.\\
\hline
    \quad   {\bf Step 1.} For each $x_i$, compute $S(x_i)$. \\
    \quad   {\bf Step 2.} For each $F_j$, compute $T(F_j)/D$.\\
    \quad   {\bf Step 3.} Return the set(s) $F_j$ with the highest
score $T(F_j)/D$.\\  \hline
\end{tabular}
\end{alg}
\smallskip

The user may now make a further selection to obtain a single
model, based either on additional information about the network to
be modeled or other criteria, such as choosing the simplest model.
This is typically a heuristic process. Alternatively one may use
this information to obtain appropriate additional data points for
further model selection. For applications to biological systems,
in particular biochemical networks, certain molecules may have
well-understood properties, such as their role in a signalling
pathway or in transcription regulation.  This type of information
can be used in the model selection process, and is generated by
the following modification of Algorithm \ref{modelselectionalg}.

\medskip
\begin{alg}
\label{modelselectionalg-bio}
  \rm
(\textbf{Model Selection-Additional Information})

\begin{tabular}{l}
    \hline
    \textbf{Input}:
    \quad A collection of subsets $F_1,\ldots ,F_t$ of $\{x_1, \ldots ,x_n\}$.\\
    \textbf{Output}:  Return all singleton sets and\\
    \quad \quad the variable(s) and set(s) of highest score.\\
\hline
    \quad   {\bf Step 1.} For each $x_i$, compute $S(x_i)$. \\
    \quad   {\bf Step 2.} For each $F_j$, compute $T(F_j)/D$.\\
    \quad   {\bf Step 3.} Return the set(s) $F_j$ with the highest
probability $T(F_j)/D$\\
\qquad \qquad \quad and all variables that have scores equal to
or higher than \\
\qquad \qquad \quad the lowest score of variables that appear in
the $F_j$\\
\qquad \qquad \quad chosen by Algorithm \ref{modelselectionalg}.\\
    \hline
\end{tabular}
\end{alg}
\smallskip


\section{Network reconstruction}
\label{net-recon}

As described in Section \ref{rev-eng}, the algorithm in \cite{LS}
chooses a reduced polynomial dynamical system that fits the given
data set. The main advantage of the algorithm described in Section
\ref{algo} is that it allows the restriction of the model space
from which this choice is made to polynomials that include only
essential collections of variables.  This improves model selection
substantially.  We demonstrate this improvement with a simulated
biochemical network in the fruit fly \emph{D. melanogaster}.

\subsection{Network reconstruction using multiple term orders}

Let $f=(f_1,\ldots ,f_{21})$ be the PDS with coordinate functions
in $\mathbb{F}_2[x_1,\ldots ,x_{21}]$ defined in the appendix.
This dynamical system was first introduced in \cite{AO} as a
Boolean network model for the segment polarity genes expressed in
a developmental cycle of the fruit fly embryo.  There the authors
assembled the Boolean functions from the known connectivity
structure, depicted as a graph in \cite[Fig. 1]{AO}. The authors
in \cite{LS} aimed to reconstruct the Boolean model, as well as
the connectivity graph, from data generated by $f$ (see Section
\ref{minset}). They applied the reverse-engineering method
described above to the generated data and constructed a minimal
PDS.  (For reasons outside of the scope of this discussion, we
focus on the reconstruction of the first 15 functions; the
remaining are associated to ``dummy'' variables introduced by the
authors of \cite{LS} and are not considered here.) To test the
accuracy of their polynomial model, they associated a directed
graph to the PDS, which they used to compare with the wiring
diagram of the Boolean model.

\begin{defn}
    Let $g\in k[x_1,\ldots,x_n]$.  The \emph{support} of $g$,
    denoted $supp(g)$, is the set of variables that appear in $g$.
\end{defn}

\begin{defn}
    Let $f$ be an $n$-dimensional PDS; that is,
    $f=(f_1,\ldots ,f_n)$ and $f_i\in k[x_1,\ldots,x_n]$.
    The \emph{dependency graph} of $f$, denoted $D(f)$, is a
    directed graph $(V,E)$ with vertex set $V:=\{x_1,\ldots,x_n\}$
    and edge set
    $E:=\{(t,x_i)\mid t\in supp(f_i), i=1,\ldots ,n\}$.
\end{defn}

The terms ``wiring diagram,'' ``static model'' and ``dependency
 graph'' are different names for the same concept and will be used
 interchangeably.

In the reconstruction process, the authors of \cite{LS} used 4
graded reverse lexicographical orders (grevlex) and produced a
consensus dependency graph. They reported 46 edges in the graph of
their polynomial model, of which 37 are correct.  As the network
graph has 44 edges, their method has a false-positive rate (FPR)
of $\frac{46-37}{46}\approx 0.20$ and a false-negative rate (FNR)
of $\frac{44-37}{46}\approx 0.15$.

While the authors demonstrated favorable performance of the
reverse-engineering method, it relies heavily on the choice(s) of
term order.  In fact, if we repeat the exercise outlined above for
only one term order, say grevlex with $x_1>\cdots >x_n$, then we
get a dependency graph with 58 edges, 37 being correct, an FPR of
$\frac{58-37}{58}\approx 0.36$ and a FNR of
$\frac{44-37}{58}\approx 0.12$. In \cite{A}, this approach was
improved by using a large number of term orders, and applied to
the reverse-engineering of a protein network.

We show next that we can improve the performance of the
reverse-engineering method proposed in \cite{LS} by using the
minimal sets algorithm described in Section \ref{algo}.

\subsection{Network reconstruction using minimal variable sets}
\label{minset}

The data set consists of 24 sets of 7 state transition pairs, each
generated by applying $f$ to 24 initializations $\mathbf{s}_0$,
taken from \cite{LS}. So, each data set is comprised of pairs
$$
\begin{array}{rl}
  (\mathbf{s}_i,f(\mathbf{s}_{i+1})), & \text{ for } i=0 \\
  (f(\mathbf{s}_i),f(\mathbf{s}_{i+1})), & \text{ for } 1\leq i\leq 7.
\end{array}
$$
There are 6 different experimental conditions represented: WT =
$f$ and KO$i$ =
$f^{(i)}:=(f_1,\ldots,f_{i-1},0,f_{i+1},\ldots,f_n)$ for
$i=2,4,6,8,12$.  The condition WT represents data from the
\emph{wildtype}, that is, observations of a biological process in
its natural state. We call $f^{(i)}$ the \emph{knockout for node
i}, as it simulates the ``knocking out'' or silencing of one
biochemical, namely a gene product.

For each experimental condition, there are 4 initializations, in
which a small number of entries are set to 1 and the rest are set
to 0. For example, the third initialization in the WT experiments
has a 1 in the first, 8th, and 12th coordinates and 0s everywhere
else. The table below summarizes this information.

\bigskip
\begin{tabular}{c|c|c|c|c|c|c}
$\mathbf{s}_0$& WT  & KO2       & KO4           & KO6       & KO8 &
KO12 \\ \hline
  1 & 4,6           & 4,6       & 6             & 4         & 4,6           & 4,6\\
  2 & 8,12,20       & 8,12,20   & 8,12,20       & 8,12      & 12,20         & 8,20 \\
  3 & 1,8,12        & 1,8,12    & 1,8,12        & 1,8,12    & 1,12          & 1,8 \\
  4 & 1,2,8,12,21   & 1,8,12,21 & 1,2,8,12,21   & 1,2,8,12  & 1,2,12,21     & 1,2,8,21
\end{tabular}
\bigskip

We applied Algorithm \ref{minsetsalg} to the generated data and
computed the minimal sets for each node. We note that minimal sets
are not unique (see Example \ref{ex}).  In fact, for only 9 of the
15 functions is there a unique minimal set of variables. Let us
restrict our attention to the following coordinate functions of the
true network for which there is more than one choice:
\begin{eqnarray*}
f_8&=&(x_4 + 1)(x_{13})[(x_{11}+1)(x_{20}x_{21} + x_{20} +
x_{21})+ x_{11}],\\
f_{9} &=&(x_{19}+1)(x_8x_9x_{18} + x_8x_9 +
x_9x_{18} + x_9) + x_8,\\
f_{10} &=&f_9(x_{20}x_{21} + x_{20}+x_{21}),\\
f_{11} &=&f_9+f_{10}+1.
\end{eqnarray*}
There are 30, 19, 2, and 5 choices of minimal sets, respectively.
In each case, we chose the set that coincided with basic biological
properties of the network. To see this, consider the subgraph $G$ of
the dependency graph of $f$ generated by the support of $f_8,\ldots
,f_{11}$.
\begin{figure}[h]
  \includegraphics[width=4in]{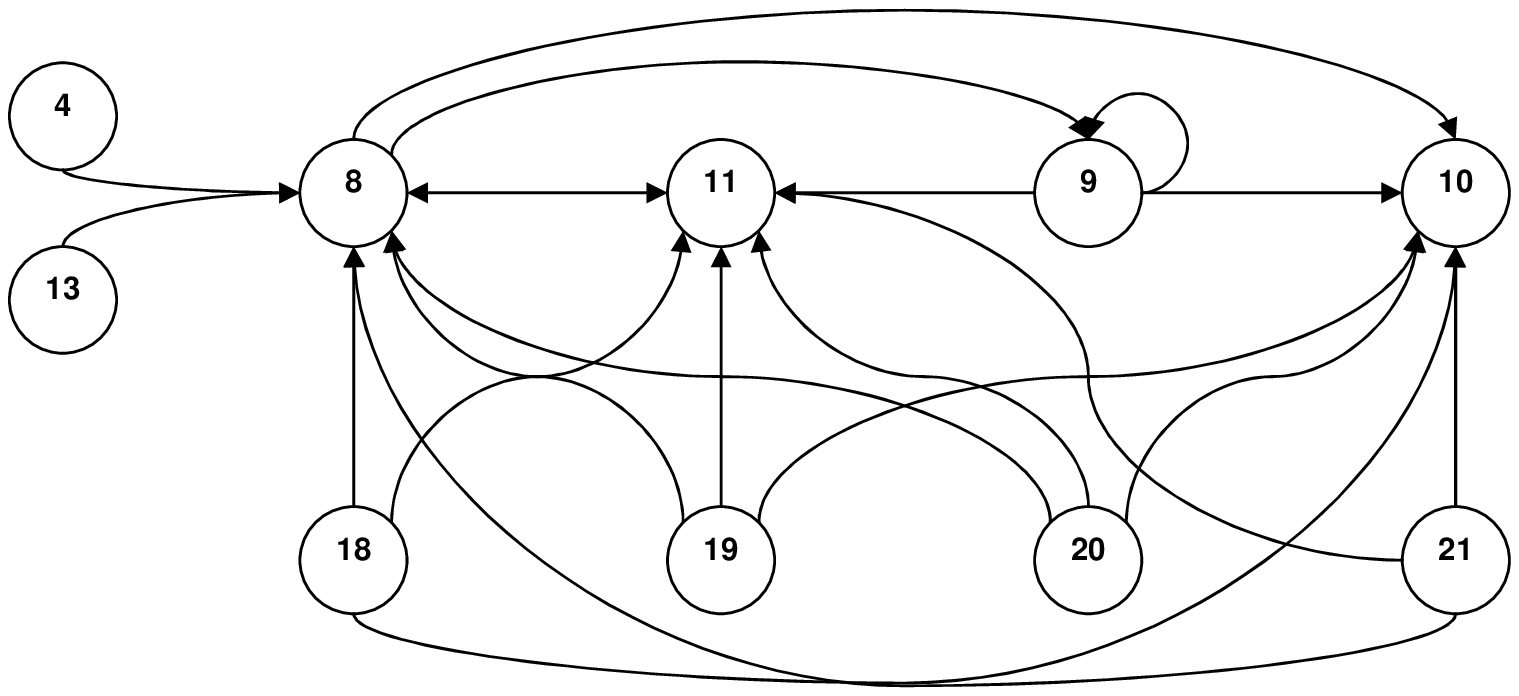}
\end{figure}
This graph has 21 edges and is given below.

The functions $f_8, f_{10}$ and $f_{11}$ are associated to
biochemicals known to not directly regulate their own synthesis,
so we selected those sets that do not contain $x_8,x_{10},$ and
$x_{11}$, respectively. We made similar selections for the
reconstruction of $f_9$.   This resulted in identification of 19
of the 21 expected edges, all of which are correct. The 2 edges
not discovered correspond to variables that are in the support of
$f_{10}$, namely $x_{18}$ and $x_{19}$. Upon inspection, we find
that the part of $f_{10}$ involving these two variables is
identically 0 on the given data set. Consequently, the
corresponding interactions in the network are not identifiable
using this data set.

For the entire network, we identified 39 edges, all of which are
correct, using Algorithms \ref{minsetsalg} and
\ref{modelselectionalg-bio}. While we failed to discover the
remaining 5 edges, all have been identified as corresponding to
elements of the ideal of the inputs.

\section{Discussion}
We have presented an algorithm that identifies all possible
minimal dependency graphs of polynomial dynamical systems that fit
a given data set of state transition pairs.  The algorithm does
not share the shortcoming of dependence on the choice of a term
order, present in the algorithm in \cite{LS}, which, on the other
hand, generates an actual dynamical system model that reproduces
the data.  And we have compared the two algorithms by applying
them to the same data set, generated from a Boolean model of fruit
fly embryonic development. Furthermore, Algorithms
\ref{minsetsalg} and \ref{modelselectionalg} improve on the greedy
algorithm described in \cite{K2}.

As with all other system identification methods of this type, a
rigorous validation requires techniques to measure the quality of
the given input data.  No such methods have been proposed at this
time for this modeling framework, an important open problem, so
validation rests on individual case studies.

\bibliographystyle{amsplain}
\bibliography{fds}

\section{Appendix}
\appendix
Following is the PDS consisting of 21 functions in
$\mathbb{F}_2[x_1,\ldots ,x_{21}]$ used as an example in Section
\ref{net-recon}.

\begin{eqnarray*}
\label{bool}
f_{1} &=&x_{1} \\
f_{2} &=&x_1x_2x_{15} + x_1x_{14}x_{15} + x_2x_{14}x_{15} +
x_1x_2 + x_1x_{14} + x_2x_{14} \\
f_{3} &=&x_{2} \\
f_{4} &=&x_1x_{16}x_{17} + x_1x_{16} + x_1x_{17} + x_{16}x_{17} + x_{16} + x_{17} \\
f_{5} &=&x_{4} \\
f_{6} &=&x_5x_{15} + x_5 \\
f_{7} &=&x_{6} \\
f_{8} &=&(x_4 + 1)(x_{13})[(x_{11}+1)(x_{20}x_{21} + x_{20} +
x_{21})+ x_{11}]\\
f_{9} &=&(x_{19}+1)(x_8x_9x_{18} + x_8x_9 +
x_9x_{18} + x_9) + x_8\\
f_{10} &=&f_9(x_{20}x_{21} + x_{20}+x_{21})\\
f_{11} &=&f_9+f_{10}+1\\
f_{12} &=&x_{5}+1 \\
f_{13} &=&x_{12} \\
f_{14} &=&(x_{11}+1)(x_{13}x_{20}x_{21} + x_{13}x_{20} +
x_{13}x_{21}+x_{13}) +
 x_{13} \\
f_{15} &=&f_{14} + x_{13}\\
f_{i}  &=&x_i\text{ for }16\leq i\leq 21
\end{eqnarray*}

\end{document}